# Fractal-like star-mesh transformations using graphene quantum Hall arrays

D. S. Scaletta,[1] S. M. Mhatre,[2] N. T. M. Tran,[2,3] C. H. Yang,[2,4] H. M. Hill,[2] Y. Yang,[5] L. Meng,[5] A. R. Panna,[2] S. U. Payagala,[2] R. E. Elmquist,[2] D. G. Jarrett,[2] D. B. Newell,[2] and A. F. Rigosi[2,a)]

[1]*Department of Physics, Mount San Jacinto College, Menifee, California 92584, USA*

[2]*Physical Measurement Laboratory, National Institute of Standards and Technology (NIST), Gaithersburg, Maryland 20899, USA*

[3]*Joint Quantum Institute, University of Maryland, College Park, Maryland 20742, USA*

[4]*Graduate Institute of Applied Physics, National Taiwan University, Taipei, 10617, Taiwan*

[5]*Graphene Waves, LLC, Gaithersburg, Maryland 20878, USA*

A mathematical approach is adopted for optimizing the number of total device elements required for obtaining high effective quantized resistances in graphene-based quantum Hall array devices. This work explores an analytical extension to the use of star-mesh transformations such that fractal-like, or recursive, device designs can yield high enough resistances (like 1 EΩ, arguably the highest resistance with meaningful applicability) while still being feasible to build with modern fabrication techniques. Epitaxial graphene elements are tested, whose quantized Hall resistance at the $\nu = 2$ plateau ($R_H \approx 12906.4$ Ω) becomes the building block for larger effective, quantized resistances. It is demonstrated that, mathematically, one would not need more than 200 elements to achieve the highest pertinent resistances.

---

a) Author to whom correspondence should be addressed.  email: afr1@nist.gov

Graphene and quantum Hall fluids have been the subject of extensive research over the past decade due to its remarkable optical and electrical properties [1-5]. Epitaxial graphene (EG), grown on 4H-SiC in the case of this work, has been developed into devices for electrical metrology due to its robust quantum Hall effect (QHE) across a wide range of magnetic fields (*B*-fields). For these devices to be successfully implemented as standards, the exhibited resistance must be well-quantized [6-10]. Devices made from EG display quantized Hall resistance values of $\frac{1}{(4m+2)}\frac{h}{e^2}$, where *m* is an integer, *h* is the Planck constant, and *e* is the elementary charge. Most EG-based devices that are used as resistance standards operate at the resistance plateau formed by the $v = 2$ Landau level ($R_\text{H} = \frac{1}{2}\frac{h}{e^2} \approx 12906.4037\ \Omega$) [11-15], with other efforts using the $v = 6$ plateau [16].

This common single-value constraint severely limits the infrastructure and equipment with which one may disseminate the unit of the ohm. Two dominant types of efforts to transcend these limitations include the use of quantum Hall array resistance standards (QHARS) to link multiple Hall elements in parallel or series and the use of *p-n* junctions, with both approaches yielding resistances of $qR_\text{H}$ where *q* is a positive rational number [17-28].

One of the limiting factors in QHARS development is the total area of high-quality EG, currently limited to the centimeter scale [29]. Other device design alternatives must be explored since these growth limitations restrict the total number of feasibly attainable QHARS elements. For instance, the maximum achievable quantized resistance from having 500 elements in series is approximately 6.5 MΩ, which is much smaller than the range of resistances currently calibrated globally – up to PΩ levels in some cases [30].

Recently, EG-based QHARS devices were used in experimental configurations enabling the application of the mathematical star-mesh transformation [31]. And though that approach can scale up to higher resistances [30, 32-34], such limits of applicability of star-mesh transformations have not yet been explored. For that reason, this work explores a framework for utilizing star-mesh QHARS device designs in a recursive manner to minimize the required number of array elements for very high effective quantized resistances. Example data from QHARS devices are also shown to support the underlying principles of this work. Given that this formulation is independent of the material's properties, it may be applied to other material systems that exhibit the QHE, as well as artifact standard resistors.

Devices were prepared in the same Si sublimation procedure described in Ref. [31] and in three major steps: (1) growth, (2) fabrication, (3) post-fabrication and packaging. Grown EG films were inspected using optical and confocal laser scanning microscopy [36], followed by fabrication of device contacts composed of NbTiN [29, 37]. Gateless control of the EG electron density for some devices was also implemented via functionalization with $Cr(CO)_3$ [38-40]. Devices were measured in a cryostat at 2 K with a Dual Source Bridge (DSB) [31].

Some of the fundamentals of star-mesh transformations are well-summarized in early work [41-42], and the framework presented herein begins by inspecting two resistance networks containing $N$ terminals, like those shown in left and right columns of Fig. 1. One may then derive a mathematical relationship between a star network (where all arms meet at a central node as in the left column of Fig. 1) and its equivalent mesh network ($N$ is equal in both networks, but the mesh contains one fewer node):

$$R_{ij} = R_i R_j \sum_{\alpha=i}^{N} \frac{1}{R_\alpha}$$

(1)

In Eq. 1, the indices go as high as $N$ with the condition that $i \neq j$. To simplify how a QHARS device undergoes minimal-element design optimization, let us define $q \equiv \frac{R}{R_H}$, where $q$ is defined as the number of single Hall elements held at the $\nu = 2$ plateau to obtain the total resistance $R$. This number has been characterized as the *coefficient of effective resistance* (CER) in other work related to graphene-based devices [43-44], with the key difference being that, for this work, functions of $q$ may be presented as analytical, wherein such functions will be easier to manage as mathematical objects than a discretized set. It should be noted that this coefficient $q$, when applied to the development of actual device designs, must be restricted to the set of positive integers ($q: q \in \mathbb{Z}^+$).

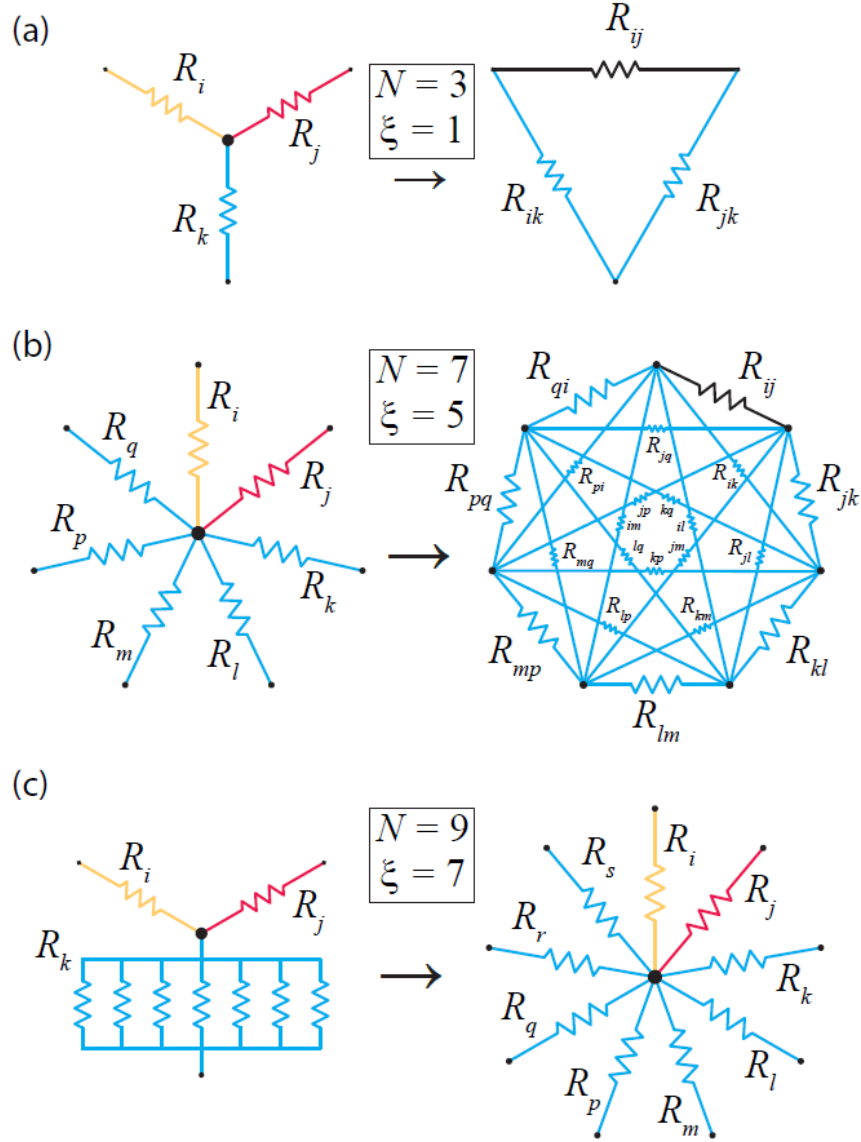

FIG. 1. Various star-mesh transformations are shown with the stars on the left and the meshes on the right (except for (c)), with $N$ and $\xi$ defined as the number of terminals and number of electrically grounded branches, respectively. (a) A Y-$\Delta$ transformation ($N = 3$). (b) A star-mesh transformation ($N = 9$). (c) Equivalence between a Y network containing a parallel set of electrically grounded elements and a star network with the same number of independent, single-element branches as there are parallel elements. The yellow and red branches represent the two primary terminals for measuring high quantized resistances.

In an experimental context, measurements are generally performed across two of the branches of a Y-$\Delta$ network, with the third branch being electrically grounded [31]. Since experimental setups only employ a high and low voltage terminal (two terminals), a reasonable condition to introduce is that all other existing terminals be grounded since future measurements of such networks would require this. Illustrations of various star-mesh transformations are shown in Fig. 1, where one defines $\xi$ as the number of grounded branches. Though this number is always two less than $N$, it is still useful to designate for subsequent mathematical manipulation. It should be noted that a Y network containing a parallel set of grounded elements is topologically equivalent to a star network with the same number of independent, single-element branches as there are parallel elements, where the ξ branches are all grounded to the same point. Therefore, the lowest number of elements any branch may contain is 1. The final two quantities to define are $D_T$, or total number of elements in a QHARS device, and $M$, a recursion factor that will be detailed later. The objective of optimization is to substantially reduce the total required elements to develop QHARS devices with quantized resistances well beyond the MΩ level. With the earlier definition normalizing quantized resistances, one may rewrite Eq. (1) as the following expression:

$$q_{ij} = q_i q_j \sum_{\alpha=i}^{N} \frac{1}{q_\alpha}$$

(2)

Next, focusing exclusively on the Y-Δ configuration for simplicity (where ξ = 1), the number of actual elements ($q$ with a single index) required to enable the measurement of a much larger, "effective", number of elements ($q$ with two indices) yields the expression of the abstract quantity $q_{ij}$ (recall from earlier that mesh resistors are virtual, not physical):

$$q_{ij} = q_i + q_j + \frac{q_i q_j}{q_k}$$

(3)

In order to maximize the effective CER in Eq. 3 ($q_{ij}$), let us impose the previously held condition that $q_k = 1$:

$$q_{ij} = q_i + q_j + q_i q_j$$

(4)

Since Eq. 4 can yield many high CERs due to the multiplication term, one can optimize device designs by finding the global minimum of this function of $q_i$ and $q_j$ that yields the desired $q_{ij}$ (which is treated as a constant, to be selected by the designer). This minimization problem may be solved with straightforward substitution and derivatives, where one temporarily defines the sum of the QHR elements in the two relevant branches to be $\alpha \equiv q_i + q_j$ (one need not include $q_k$, which has already been set to 1). Rewriting Eq. 4 in terms of $\alpha$ and $q_i$ (the latter being arbitrarily selected), one gets:

$$\alpha = \frac{q_{ij} + q_i^2}{q_i + 1}$$

(5)

And through conventional extraction of neighborhood extrema:

$$\frac{d\alpha}{dq_i} = \frac{q_i^2 + 2q_i - q_{ij}}{(q_i + 1)^2} = 0$$

(6)

The positive root of Eq. 6 yields:

$$q_i = \sqrt{q_{ij} + 1} - 1$$

(7)

This solution for $q_i$ also applies to $q_j$ given the symmetry of Eq. 4. One final check to this analysis may be computed and verified: $\frac{d^2\alpha}{dq_i^2} > 0$. This second derivative result verifies that Eq. 7 is, in fact, a global minimum and not a global maximum (global rather than local since the domain of $q$ is non-negative). To extend this analysis to the general star-mesh case, suppose there are $N$ terminals:

$$q_{ij} = q_i + q_j + \frac{q_i q_j}{q_k} + \frac{q_i q_j}{q_l} + \frac{q_i q_j}{q_m} + \cdots \frac{q_i q_j}{q_N}$$

(8)

And again, the $\xi$ branches have $q_k = q_l = q_m = \cdots = q_N = 1$:

$$q_{ij} = q_i + q_j + \xi q_i q_j$$

(9)

Equation 9, where $N = \xi + 2$, appears similar to Eq. 4, and by repeating the previous procedure:

$$q_i = \frac{1}{\xi}\sqrt{\xi q_{ij} + 1} - 1 = \frac{1}{N-2}\sqrt{(N-2)q_{ij} + 1} - 1$$

(10)

Equation 10 shows the generalized case solution for determining the CER where $N$ terminals are used. The total number of elements in the entire device may then be written: $D_T = 2q_i + \xi$. With this analysis, one may use example values of $R_{ij}$ (1 E$\Omega$, 1 P$\Omega$, 1 T$\Omega$, 1 G$\Omega$) to calculate the minimum number of required elements for those values, as in Fig. 2. Solutions ($q_i$) found for each value must be rounded to the nearest integer before calculating $D_T$. And the error introduced from rounding may be represented as deviations from the nominal example values (specifically, as $Dev = \frac{q_{ij}^{(rounded)}}{q_{ij}^{(nominal)}} - 1$). Both the deviations and $D_T$ are plotted as a function of $\xi$ in Fig. 2.

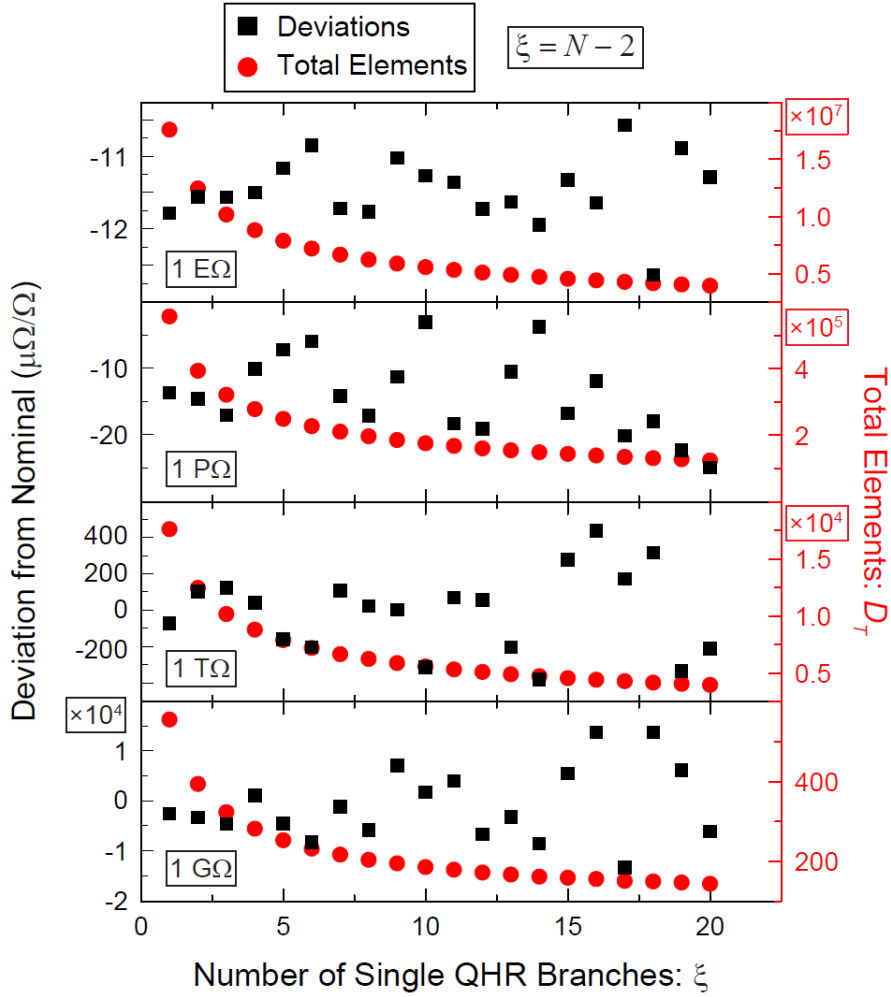

FIG. 2. Example values of high effective resistances are used to calculate the minimum number of required elements for such values (red, right side vertical axis) as well as the deviations of those hypothetical QHARS devices from the nominal example values (1 EΩ, 1 PΩ, 1 TΩ, 1 GΩ – black, left side vertical axis). The horizontal axis counts the number of single QHR branches (ξ) as a proxy for the star-mesh configuration used in the calculation.

The calculated deviations do not prevent these hypothetical QHARS devices from being useful, as many specialized Wheatstone bridges can operate without an exact decade value of resistance [31]. Though this optimization process reduced $D_T$ for GΩ level resistances to reasonable numbers for fabrication, those numbers are still high for larger resistors (in the millions for 1 EΩ). Therefore, one may expand on this optimization process by adopting recursive star-mesh designs that resemble the construction of some kinds of fractals.

Implementing a recursive treatment to star-mesh QHARS device designs may vastly expand the availability of quantized resistances. For instance, Fig. 3 summarizes a few cases in which recursive features have been included, such as embedding Y-Δ networks within Y-Δ networks. A new parameter can be designated to characterize this recursion: $M$. With a desired resistance selected, one may expand $q_{ij}$ into a star network (with all ξ branches valued as $q = 1$ and grounded). Every subsequent expansion of all non-grounded elements increases the characteristic recursion factor $M$ by one. For Fig. 3 (a)-(d), the Y-Δ networks is analyzed, whereas in Fig. 3 (e)-(h), more complex cases like the 4-terminal ($N$ = 4, or equivalently $\xi$ = 2, as defined earlier) and the 7-terminal ($N$ = 7, or equivalently $\xi$ = 5) configurations are analyzed. The complexity of the latter two cases warrants a change in representation to topologically equivalent, more abstract diagrams of device configurations.

To account for the addition of this new parameter $M$, the subscripts will be modified so as to not alter previously adopted notation; that is, actual elements represented by $q_{M:i}$ (single index) and effective number of elements represented by $q_{M:ij}$ (two indices). By repeating the optimization process in the previous section with all intermediate resistors fully expanded, leaving the QHARS device in its final configuration of elements, one obtains the actual number of elements needed per non-grounded sub-branch:

$$q_{M:i} = \frac{1}{\xi}(\xi q_{M:ij} + 1)^{2^{-M}} - \frac{1}{\xi}$$

(11)

With this information, one can count the total number of elements in the final QHARS device:

$$D_T(M, \xi, q_{M:ij}) = 2^M q_{M:i} + \sum_{x=1}^{M} 2^{x-1}\xi = \frac{2^M}{\xi}(\xi q_{M:ij} + 1)^{2^{-M}} - \frac{2^M}{\xi} + (2^M - 1)\xi$$

(12)

This function of three variables may now serve as the starting point for a final optimization process.

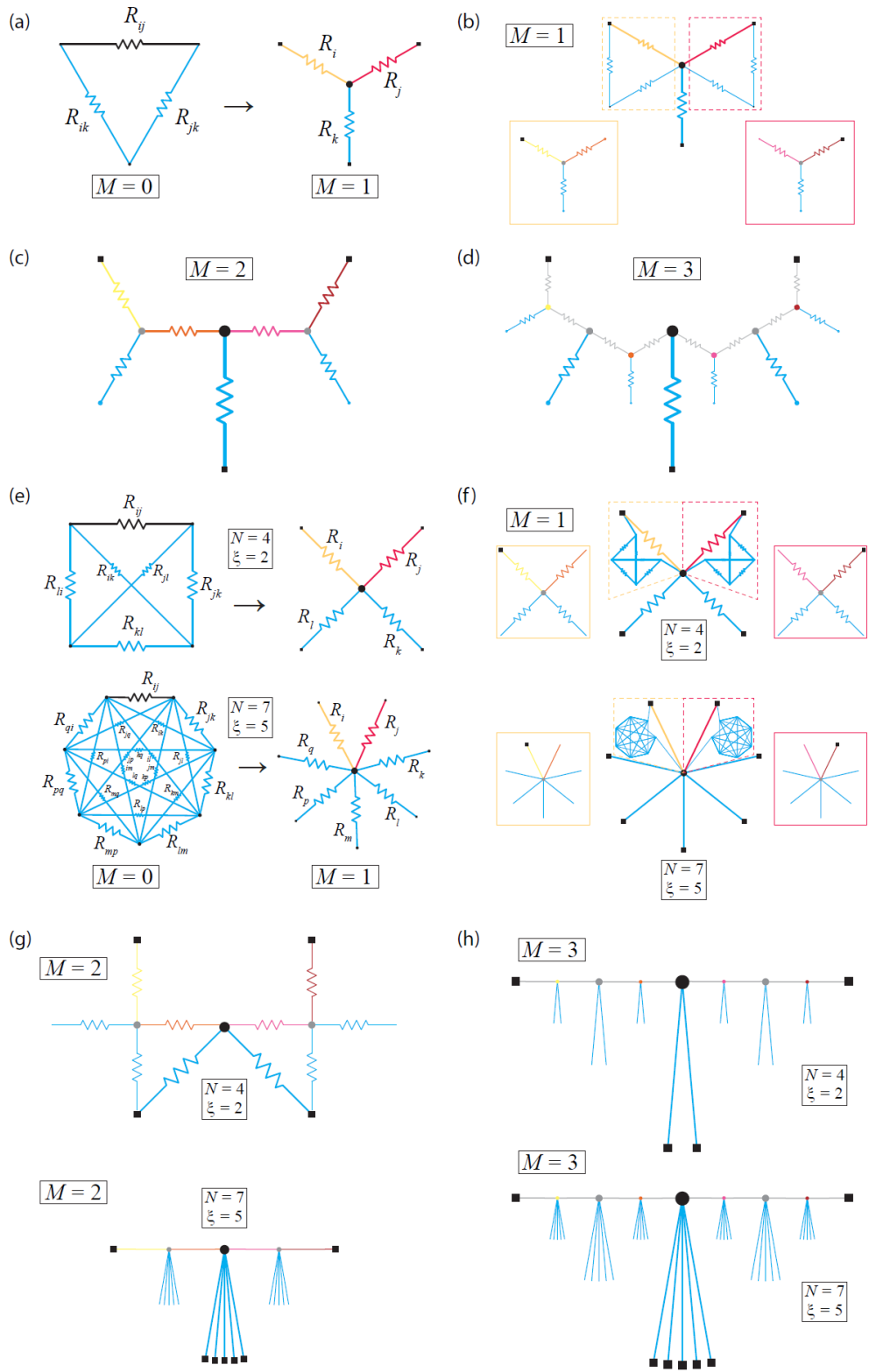

FIG. 3. (a) $R_{ij}$ (equivalently, $q_{ij}$) as a Y-Δ network. Every subsequent expansion of all non-grounded elements increases the characteristic recursion factor $M$ by one. (b) Every resistor in the non-grounded path is expanded as a Y-Δ network, with the next substitution shown in (c). (d) A third recursion is applied to all non-grounded resistors. (e) A diagram similar to (a) is produced for 4-terminal (or $N = 4$, equivalently represented in the derivation as $\xi = 2$, the number of grounded branches) and 7-terminal (or $N = 7$, equivalently represented in the derivation as $\xi = 5$, the number of grounded branches) star-mesh networks. (f) Like (b), every resistor in the non-grounded path is expanded as a network of the same number of terminals, with the next substitution shown in (g). (g) Like (c), this diagram shows the next

iteration ($M = 2$) of the recursion for both the 4-terminal (top) and 7-terminal (bottom) cases. Note that the top and bottom appear different in drawing format. This topologically equivalent diagram of the device configurations must be used due to the added difficulty of drawing more complex meshes with each recursive iteration. (h) Like (d), this diagram shows the next iteration ($M = 3$) of the recursion for the two cases ($N = 4$ and $N = 7$).

As per the construction of Eq. 12, it would benefit the designer to first select a desired final element count $q_{M:ij}$ (proxy for resistance). For the sake of example, the upper bound on arguably useful resistances for metrology is selected: 1 E$\Omega$ ($q_{M:ij} \approx 7.74809 \times 10^{13}$).

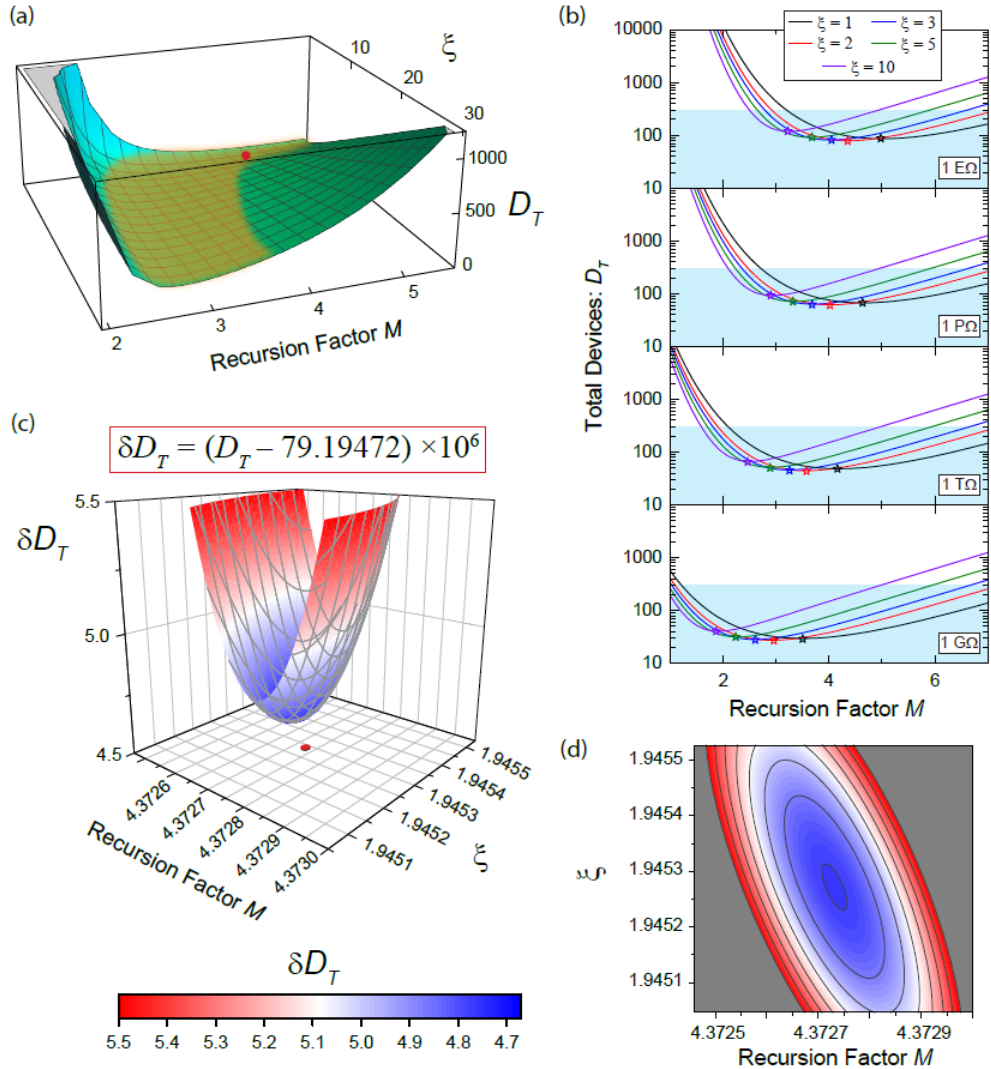

FIG. 4. (a) $D_T$ is plotted as a function of $\xi$ and $M$, while fixing $q_{M:ij}$ corresponding to 1 E$\Omega$. The minimum is marked as a red spot, and the shaded orange region refers to the embankment of practically feasible design solutions in the text. (b) The behavior of $D_T$ as a function of $M$ at various fixed values of $\xi$. $D_T$ is plotted for values of $q_{M:ij}$ corresponding to 1 P$\Omega$, 1 T$\Omega$, and 1 G$\Omega$. Shaded cyan regions indicate an upper bound of 300 elements, a rough approximation for fabrication capabilities. Every curve has its minimum marked by a star. (c) Magnification of global minimum of $D_T$ according to the listed transformation. A corresponding contour projection is shown in (d).

Attempting to analytically find the minimum of this now two-variable function quickly reveals that one must solve a large order polynomial, and since such large polynomials are not guaranteed to have closed-form solutions (see SM [45]), finding the global minimum must be done numerically. Fixing $q_{M:ij}$ to correspond to 1 E$\Omega$, one may plot this function as seen in Fig. 4 (a). The global minimum is marked as a red spot and can be greatly magnified to see the subtle approach to that extremum (see Fig. 4 (c) and (d)). Inspecting this solution space yields two observations. First, there is some flexibility in how to design a device since there are a set of solutions with low $D_T$ along the embankment region of $M = 3$ and $\xi = 2$ to 4 (shaded orange in Fig. 4 (a)). The embankment is also presented in Fig. 4 (b), where semi-logarithmic curves for $D_T$ are plotted for fixed values of $\xi$ (and for 1 E$\Omega$, 1 P$\Omega$, 1 T$\Omega$, and 1 G$\Omega$). Shaded cyan regions indicate an upper bound of 300 elements, which is a rough approximation for what is likely within fabrication capabilities [28]. The reason one should take note of this cyan region is because exclusively using the global minimum, when accounting for fact that one must use integers in practice, may not give a nominal resistance value close enough to the desired value, as defined by the experimenter (see SM for example calculations).

Generally, optimal configurations using high recursion make it inflexible to achieve an exact desired value with a low error. What can be learned by calculating example device designs is that the likelihood of obtaining a resistance close to a desired value is statistically greater when $M$ is taken to be 2 or 3 since the embankment region in Fig. 4 (a) allows for greater flexibility in $\xi$ (and thus more chances at obtaining a combination of parameters yielding an optimally accurate resistance). See the SM for more details [45].

With greater flexibility in QHARS designing, one may now more closely analyze the extent to which such flexibility can improve nominal value accuracy. For this analysis, $M = 3$ (initially) since $M = 4$ too greatly restricts the parameter space. With a chosen $q_{3:ij}$, $D_T$ can be minimized in terms of $\xi$. This minimization can be seen in Fig. 5 (a) on the red vertical axis ($D_T$). Naturally, one selects an integer $\xi$ such that $D_T$ is minimized and proceeds to obtain $q_{3:i}$. As done in the SM [45], deviations of the final device design may be calculated (represented as black squares in Fig. 5 (a)). Furthermore, rounding to the closest integer may not always yield the optimal deviation, as seen in the 1 G$\Omega$ case, where the optimal $\xi$ yields a resistance, whose deviation falls off-scale. Therefore, it may be fitting to perturb the solution of integer parameters to see if more optimal solutions exist. Such perturbations are also shown in Fig. 5 (a) as black hollow triangles, indicating that better accuracies are available at limited increases in $D_T$. These more optimal parameters are rendered into QHARS device designs in the SM [45].

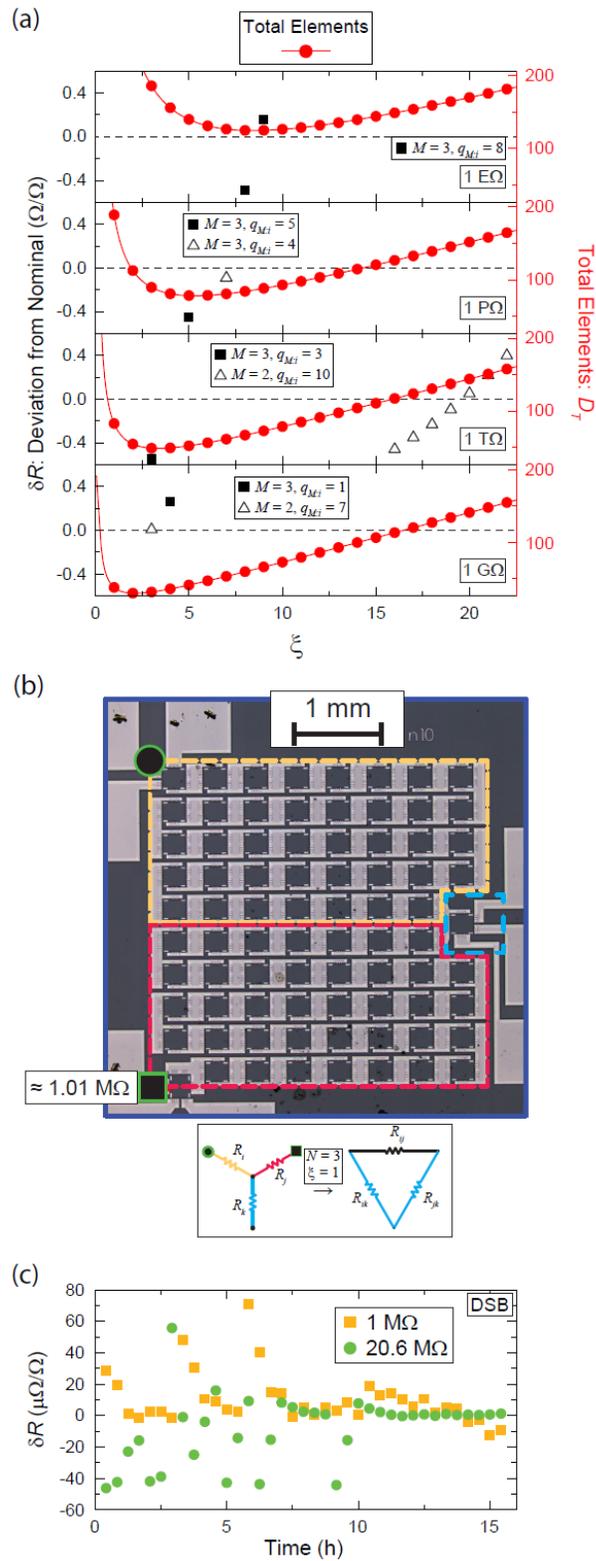

FIG. 5. (a) With $q_{3:ij}$ chosen, $D_T$ can be minimized in terms of $\xi$ on the red vertical axis. Deviations of the final device design may be calculated from the solutions closest to the minima when rounded to integers (represented as black squares). When solutions are perturbed by alternative means of rounding to an integer (like alternating rounding down and up for the relevant parameters), possible improvements may be found to the accuracy of the resistance output when compared to the desired resistance (black, hollow triangles). (b) An example image of a device valued nominally at about 1 MΩ due to its 78 devices in series, with a middle element on the right side, center, as the grounded branch. Accompanying illustration shows the corresponding diagramatic representation of the Y-Δ transformation ($N = 3$). (c) Example measurement of a QHARS device using a DSB, both across its full array (gold) and in the Y-Δ configuration (green).

To test these concepts, several device measurements were performed on a symmetric Y-Δ network. A DSB was used, as in Ref. [31], to more precisely measure the output resistances of one type of QHARS (see SM [45]) in Fig. 5 (b), where the offsets from nominal for the values 1 MΩ and 20.6 MΩ were 1812.6 Ω and 6699.473 Ω, respectively. The demonstration of the device's ability to be used in a star-mesh network makes it evident that this mathematical transformation has the promise of substantially reducing the complexity of future QHARS devices, especially those in use for electrical metrology.

In conclusion, a mathematical approach was adopted for optimizing the number of total elements required for obtaining high effective quantized resistances in graphene-based quantum Hall array devices. Star-mesh transformations involving recursive device designs yielded resistances as high as nearly 1 EΩ. Furthermore, designs may be lightly modified to enhance accuracy to a desired value while still remaining feasible to build with modern fabrication techniques. A general observation based on the results suggests that using fewer recursions would allow the greatest flexibility in device design, including designs that are not wholly symmetric in their star-mesh branches as treated in this work. Lastly, these designs are encouraged to be tested for their promise in removing the need for artifact resistors and for providing access to the quantum SI without the need for a lengthy calibration chain.

## SUPPLEMENTARY MATERIAL

The Supplementary Material includes additional mathematical details, calculations for showing deviations from nominal resistances, examples of other device designs, and QHARS devices used in experiments.

## ACKNOWLEDGMENTS

The work of C.H.Y. and S.M.M. at NIST was made possible by C.-T. Liang of National Taiwan University, and the authors thank him for this endeavor. The authors would like to express thanks to T. Mai, F. Fei, G. J. Fitzpatrick, and E. C. Benck for their assistance in the NIST internal review process.

The authors declare no competing interests. Commercial equipment, instruments, and materials are identified in this paper in order to specify the experimental procedure adequately. Such identification is not intended to imply recommendation or endorsement by the National Institute of Standards and Technology or the United States government, nor is it intended to imply that the materials or equipment identified are necessarily the best available for the purpose.

# Supplementary Material

D. S. Scaletta,[1] S. M. Mhatre,[2] N. T. M. Tran,[2,3] C. H. Yang,[2,4] H. M. Hill,[2] Y. Yang,[5] L. Meng,[5] A. R. Panna,[2] S. U. Payagala,[2] R. E. Elmquist,[2] D. G. Jarrett,[2] D. B. Newell,[2] and A. F. Rigosi[2]

[1]*Department of Physics, Mount San Jacinto College, Menifee, California 92584, USA*

[2]*Physical Measurement Laboratory, National Institute of Standards and Technology (NIST), Gaithersburg, Maryland 20899, USA*

[3]*Joint Quantum Institute, University of Maryland, College Park, Maryland 20742, USA*

[4]*Graduate Institute of Applied Physics, National Taiwan University, Taipei, 10617, Taiwan*

[5]*Graphene Waves, LLC, Gaithersburg, Maryland 20878, USA*

Table of Contents:

1. Additional mathematical details
2. Calculations for showing deviations from nominal resistances
3. Examples of other device designs
4. QHARS devices used in experiments

# 1 Mathematical Details

In general, polynomials of degree $n \geq 5$ are not solvable by radicals. That is, it is not possible to solve a polynomial of degree $n \geq 5$ with arbitrary rational coefficients algebraically in closed form to derive a general solution, such as the quadratic formula in the case of polynomials of degree $n = 2$. This is a well known result in pure mathematics called the Abel-Ruffini theorem. There are special cases of polynomials that can be solved, and determining the solvability of polynomials of degree $n \geq 5$ is a large part of Galois theory, which is a field of study in abstract, or modern algebra, that seeks to reduce problems in field theory to problems in group theory. In this context, field theory is not to be mistaken for the fields studied in physics, rather, fields studied in mathematics. The most familiar fields would be the real numbers, $\mathbb{R}$, and the complex numbers, $\mathbb{C}$. In fact, when Evariste Galois (1811-1832) began his work, he was attempting to prove that the polynomial $x^5 - 6x + 3$ had no zeros.

An extremely trivial example of a degree five polynomial (quintic) that can be solved is given by $x^5 - 1 = 0$. On the other hand, the most simple quintic that cannot be solved is $x^5 - x - 1 = 0$. The proof of the Abel-Ruffini theorem predates the advent of Galois theory, but the most common and least esoteric proof uses Galois theory. Because of the pertinence of Galois theory to both proving the Abel-Ruffini theorem and determining the solvability of polynomials of degree $n \geq 5$, we will provide a rudimentary introduction here. We would like to emphasize that Galois theory is a large and complex field of study that is generally covered in the second semester of a sequence of algebra courses. It is a very powerful tool that is used throughout many fields of research in pure and applied mathematics.

The fundamental question Galois theory seeks to answer is that if we are given a polynomial,
$$p(x) = a_n x^n + a_{n-1} x^{n-1} + \cdots + a_1 x + a_0,$$
with rational coefficients, can we find its roots? That is, can we find any constants $c$ such that $p(c) = 0$? If we can find all of its roots, the polynomial factors, or *splits*, as
$$p(x) = a_n \prod_{i=1}^{n} (x - c_i).$$

So in some regard, we are simply trying to find the smallest set of numbers, or field, containing the roots with which $p(x)$ will split. This field will always be an extension of $\mathbb{Q}$ contained in the complex numbers $\mathbb{C}$. If that field exists, then we can solve the polynomial for all of its roots. If no such field exists, then the polynomial will not fully factor, and is therefore not solvable. Let us begin with some necessary definitions, examples and background.

**Definition 1.** *A group $(G, \times)$ is a set, $G$, together with an operation $\times$ such that*

1. *If $x$ and $y$ are elements of $G$, then $x \times y$ is also in $G$.*

2. *There must exist an element $1$ in $G$ such that $1 \times x = x \times 1 = x$ for every element $x$ in $G$. The element $1$ is called the identity in $G$.*

3. *For any arbitrary elements $x$, $y$ and $z$ in $G$, the group operation is associative. That is, $(x \times y) \times z = x \times (y \times z)$.*

4. *For every $x$ in $G$, there is a unique element $x^{-1}$ in $G$ called the inverse of $x$ such that $x \times x^{-1} = x^{-1} \times x = 1$.*

Just like sets, groups can be contained in larger groups, or can contain smaller groups within them.

**Definition 2.** *Given a subset $H \subset G$, we say that $H$ is a subgroup of $G$ if the operation on $G$ restricts to $H$ and the identity $1 \in G$, is the identity in $H$.*

There are many examples of groups with which everyone is familiar, such as the integers $\mathbb{Z}$ together with standard addition as its operation. In this instance the identity is 0. The integers are also a group under multiplication with identity 1. The same can be said of the real numbers $\mathbb{R}$. The aforementioned groups have infinitely many elements, but a pair of groups of particular importance to Galois theory are the cyclic group and the symmetric group, both of which are finite groups.

**Definition 3.** *The cyclic group of order $p$, denoted $C_p$, is the group with $p$ elements $1, x, x^2, x^3, ..., x^{p-1}$ together with the operation $x^q \times x^r = x^{q+r}$ and the relation that $x^p = 1$.*

**Definition 4.** *Given a set of $n$ elements, $X = 1, 2, ..., n-1, n$, the bijections $f : X \to X$ are called permutations of $X$. The set of all such bijections forms a group where the operation is the composition of these bijections. This group is called the symmetric group, denoted $S_n$.*

Symmetric groups can more simply be understood as groups of rearrangements, or permutations. We can therefore see that $S_n$ is the set of permutations of the set $1, 2, ..., n$, which again forms a group where the operation is the composition of permutation. It is easy to see that $S_n$ has $n!$ elements. As we previously mentioned, Galois theory is about reducing problems in field theory to problems in group theory, so let us introduce the concept of a field. Fields are in fact groups, but they have an extra operation as well as the necessary added structure.

**Definition 5.** *A field $(\mathbb{F}, +, \times)$ is a set, $\mathbb{F}$, together with operations $+$ and $\times$ such that*

1. *$\mathbb{F}$ is a group under the operation $+$ with identity element 0.*

2. *$\mathbb{F}$ without the element 0 is a group under the operation $\times$ with identity element 1.*

3. *Any three elements in $\mathbb{F}$ must satisfy the distributive property. That is, for any $x, y, z$ in $\mathbb{F}$, we have that $x \times (y + z) = x \times y + x \times z$.*

4. *The operation $+$ is commutative, so $x + y = y + x$ for any $x, y$ in $\mathbb{F}$.*

5. *For any $x$ in $\mathbb{F}$, we have that $x \times 0 = 0 \times x = 0$.*

The definition of a field is meant to generalize sets of numbers into an abstract concept. Notice that the sets of rational, real and complex numbers all meet this definition with common addition and subtraction. A less obvious field that is of particular importance to Galois theory is $\mathbb{Q}[\sqrt{2}] = \{a + b\sqrt{2} | a, b \in \mathbb{Q}\}$. A similar example we will use later is $\mathbb{Q}[\sqrt{2}, \sqrt{3}] = \{a + b\sqrt{2} + c\sqrt{3} + d\sqrt{6} | a, b, c, d \in \mathbb{Q}\}$. Notice that the elements of these sets generally look like solutions to polynomials, and that is the point. We will be constructing sets that are the smallest that contain the solutions particular polynomials. Before we do, we have a few more definitions to cover, the first of which will define the types of numbers with which we can build these particular sets.

**Definition 6.** *A number $\alpha$ is called algebraic if it is a zero of some nonzero polynomial.*

A number must be algebraic for us to construct one of these fields $\mathbb{Q}[\alpha]$ as described above. These fields are important, because if we are given a polynomial

$$p(x) = a_n x^n + a_{n-1} x^{n-1} + \cdots + a_1 x + a_0,$$

where the coefficients are rational numbers and $p(\alpha) = 0$, then $\mathbb{Q}[\alpha]$ is the smallest field containing $\mathbb{Q}$ that also contains $\alpha$. That is, $\mathbb{Q}[\alpha]$ extends $\mathbb{Q}$ just enough to contain the solutions of $p(x)$ which are constructed as linear combinations of $\alpha$. To put it another way, $\mathbb{Q}[\alpha]$ is the set of elements of the form $a_{n-1}\alpha^{n-1} + \cdots + a_1\alpha + a_0$ such that the coefficients are all rational numbers and $n$ is the smallest integer such that there exists a polynomial $p(x)$ of degree $n$ having a zero $\alpha$. We can easily construct examples of fields like these, for instance, $\mathbb{Q}[\sqrt[3]{3}] = \{a_0 + a_1\sqrt[3]{3} + a_2\sqrt[3]{3}^2 | a_0, a_1, a_2 \in \mathbb{Q}\}$. These extended fields are of importance to the current issue because their existence will be tantamount to being able to find the zeros of a polynomial. Let us formally define these fields.

**Definition 7.** *A field extension of $\mathbb{F}$ is a field $\mathbb{E}$ containing $\mathbb{F}$. This relationship between a field and its extension is commonly denoted $\mathbb{E}/\mathbb{F}$, or in the set theoretic notation as $\mathbb{F} \subseteq \mathbb{E}$*

**Definition 8.** *Given a polynomial $p(x)$ with rational coefficients, the splitting field of $p(x)$ is the smallest field extension of $\mathbb{Q}$ that contains all the zeros of $p(x)$.*

We have already seen some examples of splitting fields. For example, $\mathbb{Q}[\sqrt{2}]$ is the splitting field for $x^2 - 2$, and $\mathbb{Q}[\sqrt{2}, \sqrt{3}]$ is the splitting field for $x^4 + 5x^2 + 6$. An example of particular interest is the splitting field for $x^2 + 1$, which is the set of complex numbers, $\mathbb{C}$. So now we are starting to see how proving the existence of a particular field, the splitting field, is tantamount to discovering whether or not a polynomial does indeed split. Galois theory provides us with the mathematical structure necessary to construct these proofs. It is a large subject and can be quite technical, so it is not feasible to provide a complete introduction here, but there are many excellent references on the subject, from Gallian's fairly simple text to Serge Lang's graduate text *Algebra*. The goal here is to give the reader a rudimentary understanding of how the theory applies and how one uses it to determine if a polynomial indeed has zeros.

Before we continue, let us focus on some interesting properties that certain roots have. We notice that roots of even degree polynomial come in pairs. For instance, the most familiar pair of roots is given to us by none other than the quadratic formula,

$$x = \frac{-b \pm \sqrt{b^2 - 4ac}}{2a},$$

for the polynomial $ax^2 + bx + c$. We can also look at $x^4 + 5x^2 + 6$, which factors as $(x+\sqrt{2})(x-\sqrt{2})(x+\sqrt{3})(x-\sqrt{3})$, so we again have pairs of solution, $x = \pm\sqrt{2}, \pm\sqrt{3}$. What we notice about these pairs of roots is that they are conjugates of one another. So we should formally define such a map in order to utilize this symmetry. In particular, we want a map that would take an element in a splitting field to it's conjugate. For example, $f : \mathbb{Q}[\sqrt{3}] \to \mathbb{Q}[\sqrt{3}]$ where $f(a+b\sqrt{3}) = a - b\sqrt{3}$, or $f : \mathbb{C} \to \mathbb{C}$ such that $f(a+bi) = a - bi$. Such a map is called a field automorphism.

**Definition 9.** *A field automorphism is a bijective map $f$ that takes elements of $\mathbb{F}$ to elements of $\mathbb{F}$, $f : \mathbb{F} \to \mathbb{F}$ satisfying that for any $x, y \in \mathbb{F}$, we have*

1. $f(x+y) = f(x) + f(y)$,

2. $f(x \times y) = f(x) \times f(y)$,

3. $f(1/x) = 1/f(x)$.

Field automorphisms are maps from the field back into itself that preserve the algebraic structure of the field. We need to extend this definition a bit in order to make it useful. In fact, we will construct the central component of Galois theory out of the following objects

**Definition 10.** *Given a field extension $\mathbb{E}/\mathbb{F}$, then an $\mathbb{F}$-automorphism of $\mathbb{E}$ is an automorphism of $\mathbb{E}$ that fixes $\mathbb{F}$. That is, for any element $x \in \mathbb{F}$, $f(x) = x$.*

Notice how the $\mathbb{F}$-automorphism of $\mathbb{E}$ operation preserves the algebraic structure in the same way as conjugation. In particular, we can see that just like with conjugation, if $p(x)$ is a polynomial with coefficients in $\mathbb{F}$ then we have $f(p(x)) = p(f(x))$. So these functions preserve the algebraic structure of $\mathbb{E}$ and therefore $\mathbb{F}$, leave elements of $\mathbb{F}$ unchanged, and effectively permute elements of the extension $\mathbb{E}$ that are not in $\mathbb{F}$. There are usually not very many of these automorphisms in many cases, and you can use the properties of field automorphisms in order to determine the possibilities. For example, if we take a look at the $\mathbb{Q}[\sqrt{3}]$ automorphisms of $\mathbb{Q}$ the properties above require that one of them must simply be the identity function on $\mathbb{Q}$. But if $f$ is such an automorphism and is applied to $\mathbb{Q}[\sqrt{3}]$ it must satisfy the fact that

$$3 = f(3) = f(\sqrt{3}\sqrt{3}) = f(\sqrt{3})f(\sqrt{3}) = f(\sqrt{3})^2.$$

This shows us that $f(\sqrt{3}) = \pm\sqrt{3}$, so there are only two $\mathbb{Q}[\sqrt{3}]$ automorphisms of $\mathbb{Q}$. These are the identity map and the map $f(a + b\sqrt{3}) = a - b\sqrt{3}$. We can see that f is actually its own inverse, since $f(f(a + b\sqrt{3})) = f(a - b\sqrt{3}) = a + b\sqrt{3}$ or $f$ is the identity, which is trivially its own inverse. Furthermore, the set of all the $\mathbb{Q}[\sqrt{3}]$ automorphisms of $\mathbb{Q}$ forms a group under the operation of composition. In fact, this is something we can say in general!

Now, recall that these functions either fix elements in $\mathbb{E}$ or permute them. This should remind us of the symmetric group, and indeed it is no coincidence.

**Definition 11.** *Given a field extension $\mathbb{E}/\mathbb{F}$, the set of all $\mathbb{F}$-automorphisms of $\mathbb{E}$ forms a group, called the Galois group of $\mathbb{E}$ over $\mathbb{F}$, denoted $Gal(\mathbb{E}/\mathbb{F})$ or $Aut_{\mathbb{F}}(\mathbb{E})$.*

**Definition 12.** *If $f$ is a polynomial with rational coefficients and $\mathbb{E}$ is its splitting field over $\mathbb{Q}$, then $Gal(\mathbb{E}/\mathbb{Q})$ the Galois group of the polynomial.*

This is the point where we are going to have to discard any semblance of rigor from the perspective of the pure mathematician, which is sufficient here because we only seek to give the flavor of the underlying theory behind our statement about the insolvability of polynomials by radicals. There is a crucial theorem at the heart of Galois theory called the fundamental theorem of Galois theory that provides us with the ever so important relationship between fields and groups. In particular, it states that there is a one to one correspondence between the splitting fields of polynomials and their respective Galois groups. There a number of conditions that these correspondences have, and mathematicians have ways of diagramming lattices of these subgroups that elucidate the configuration by which the splitting fields are contained in each other.

There is a notion of solvability in group theory, which loosely stated means that given a group $G$, we say it is solvable if there is a sequence of subgroups $G = G_0 \supset G_1 \supset \cdots \supset G_k = \{1\}$ satisfying particular conditions between the containments $G_i \subset G_{i-1}$. The word solvable refers to the solvability of polynomials. A crucial result discovered by Galois in 1830 is the fact that if $f$ is a polynomial with Galois group $Gal(\mathbb{E}/\mathbb{Q})$, then if $f$ is solvable, then $Gal(\mathbb{E}/\mathbb{Q})$ is a solvable group. This, of course, implies that if the Galois group is not solvable, then $f$ is not solvable by radicals. As an example of how this is used, it can be shown that $2x^5 - 10x + 5$ has Galois group $S_5$, which is known to be unsolvable, therefore so is the polynomial.

It is important to note that when polynomials can be determined to not be able to be factored using Galois theory, that there are a couple of possibilities. The first case is that the polynomial simply has no zeros. The second is that the polynomial might have zeros, but not as many zeros as its degree. This does not mean a polynomial has zeros of multiplicity greater that one. A polynomial still splits if it has zeros of multiplicity greater than unity, such as $(x-1)^3$. This is indeed a polynomial of degree $n = 3$ having three zeros, but each zero is given by unity, so we say it has the zero $x = 1$ of multiplicity three. What we're referring to is a polynomial of degree $n$ such as

$$f(x) = g(x) \prod_{i=1}^{k}(x - a_i),$$

where $g(x)$ is an insolvable polynomial of degree $n - k$. In the second case, and even in the case of a polynomial that splits completely, we can still find the zeros by certain approximation techniques. We know that there is no general formula for solving a polynomial of degree $n \geq 5$, but we still have techniques for finding zeros. For example, we can hone in on zeros by using Newton's method. This is completed by approximating the zeros to the best of our ability. Once we find a value $x_0$ so that $f(x_0)$ is reasonably close to zero, we then find the tangent to $f(x)$ at $x_0$ and follow it back to the $x$-axis where we get a slightly closer value to the actual zero, given by

$$x_1 = x_0 - \frac{f(x_0)}{f'(x_0)},$$

and keep iterating as

$$x_n = x_{n-1} - \frac{f(x_{n-1})}{f'(x_{n-1})}.$$

In as few as four iterations, you can have a zero approximated to as many as fifteen decimal places.

Recursive star-mesh configurations appear to have properties quite similar to fractals. These are what mathematicians call pre-fractals. Fractals exhibit two key properties, the first being that they are recursive and the second is that they are self similar. The first property will generally be given by some iterated sequence of functions. In this way, a fractal can actually be viewed as the orbit, or locus of points related by the evolution function of a dynamical system. Fractals can also be constructed topologically. For example, the Cantor set is the fractal constructed taking the line segment $[0, 1]$ and removing the middle third, then removing the middle third of the two remaining pieces, then removing the middle third of the four remaining pieces, ad infinitum. The second property is that each piece of the fractal looks similar to its whole self. If you were to keep zooming in and looking at smaller and smaller regions of the fractal, you will see the same picture over and over. Fractal like structures are ubiquitous in nature. They can be observed in the structure of trees, to the way species distribute their populations, to the structure of the axons connecting our neurons in our brains.

## 2. Calculations for Showing Deviations from Nominal Resistances

A useful algorithm may be devised to very roughly guide how one should design devices based on the stringency of the desired value.

To begin, let us consider the global minimum found numerically in the main text for $q_{M:ij}$ fixed to correspond to 1 E$\Omega$ (Fig. 4 (a)). The numerical solution indicates that $M = 4.37273$ and $\xi = 1.94528$ (yielding a minimum of about 79.19 elements). Obviously, as one applies the condition that real devices require all quantities to be integers, one must approximate the ideal set of parameters by simple rounding. Doing so would then require a calculation of two more parameters, namely $q_{M:i}$ (which, recall, is the number of elements for each branch of a fully expanded star-mesh recursion design) and $D_T$, which should be the final calculated quantity since we are now placing weight on the importance of limiting the deviation from a desired value (and such a concession may cost additional elements). We have:

$$q_{M:i} = \frac{1}{\xi}(\xi q_{M:ij} + 1)^{2^{-M}} - \frac{1}{\xi}$$

(S2)

Eq. S2 (Eq. 11 in the main text), now having approximated $M$ and $\xi$ to integers, will yield a value for $q_{M:i}$ that will very likely not be an integer (about 3.3 in the exemplary case above). The next step is to solve for the new final, large CER $q_{M:ij}$, after having modified several variables, by means of inverting Eq. S2:

$$q_{M:ij} = \frac{(\xi q_{M:i} + 1)^{2^M} - 1}{\xi}$$

(S3)

When $q_{M:ij}$ is calculated with the three integers, one obtains a final CER of about $1.66 \times 10^{13}$ elements, corresponding to an approximate 0.2 E$\Omega$. The deviation may be calculated as in the main text: $Dev = \frac{q_{ij}^{(rounded)}}{q_{ij}^{(nominal)}} - 1$. If one inspects Fig. 4 (a) in the main text, it is clear that there is some flexibility in how to design a device since there are a set of solutions with low $D_T$ along the embankment region (shaded orange) of $M = 3$ and $\xi = 2$ to 4 and beyond. If one wishes to improve on the accuracy of a final device design vis-à-vis customized, selected resistance, then additional strategies may include rounding integers up or down in alternating fashion (rather than "closest integer") as though to compensate for the reduction or increase in expected deviations.

This strategy is employed for four examples of devices, which had first been optimized for $D_T$, but from which minimum the three parameters $M$, $\xi$, and $q_{M:i}$ were adjusted slightly within their numerical neighborhoods and in a trial-by-error fashion. This approach yields a set of more optimal devices (as displayed in Fig. 5 in the main text) when using the metric of its resistance being close to the desired value.

## 3. Examples of Other Device Designs

Though the following designs have been drawn out to exemplify the extent to which $D_T$ may be minimized, they should not be taken to be the optimal designs for very specifically customized values of resistance. Such a design must take into consideration a balance of $D_T$ and desired deviation from nominal, as described in the previous section. For the values 1 E$\Omega$, 1 P$\Omega$, 1 T$\Omega$, and 1 G$\Omega$, several device designs are displayed below as more optimal than their constructions derived from simply rounding each parameter (from the solution of the $D_T$ global minimum) to the nearest integer. As stated in the main text, one may include additional devices in an attempt to seek a more accurate value to one's desired resistance. It should be noted that in the case of Fig. 4-SM ($M = 2$ and $\xi = 3$, $q_{M:i} = 7$, and $D_T = 37$), an artifact-based resistance network was built using 90.9 k$\Omega$ (approximately $q_{M:i} = 7$) and 4.32 k$\Omega$ (approximately $\xi = 3$) resistors. This network was then measured with a commercial teraohmmeter at 50 V, yielding about 1.02 G$\Omega$.

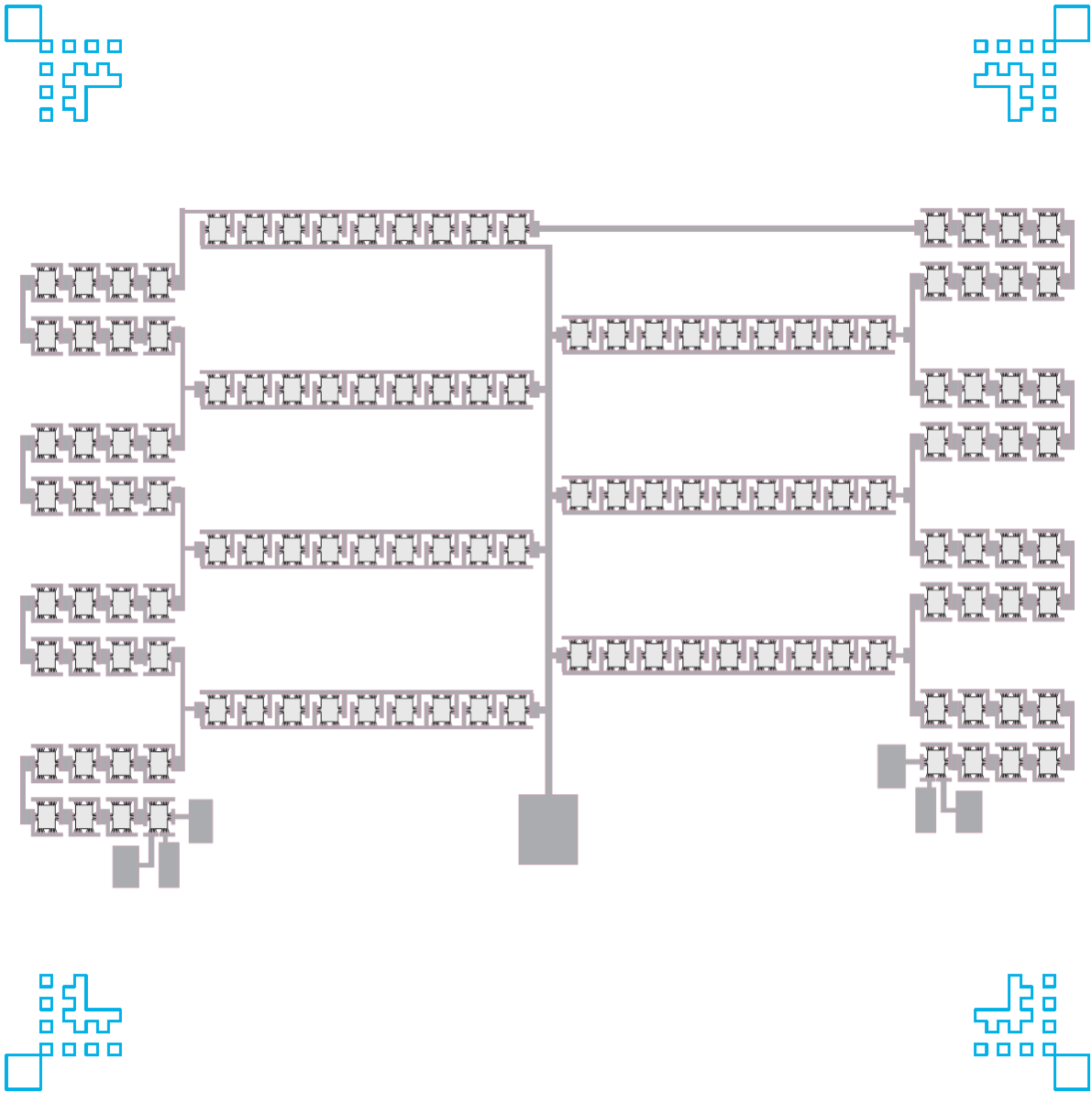

FIG. 1-SM. A design layout is shown for the following parameters: $M = 3$ and $\xi = 9$, $q_{M:i} = 8$, and $D_T = 127$. (Recall that $\xi$ represents the single QHR elements in parallel). The desired value was 1 E$\Omega$, and this arrangement reaches within approximately 85 % of that nominal value (that is, nearly 0.85 E$\Omega$).

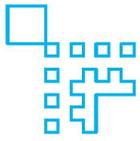
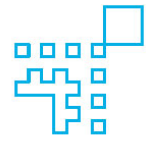
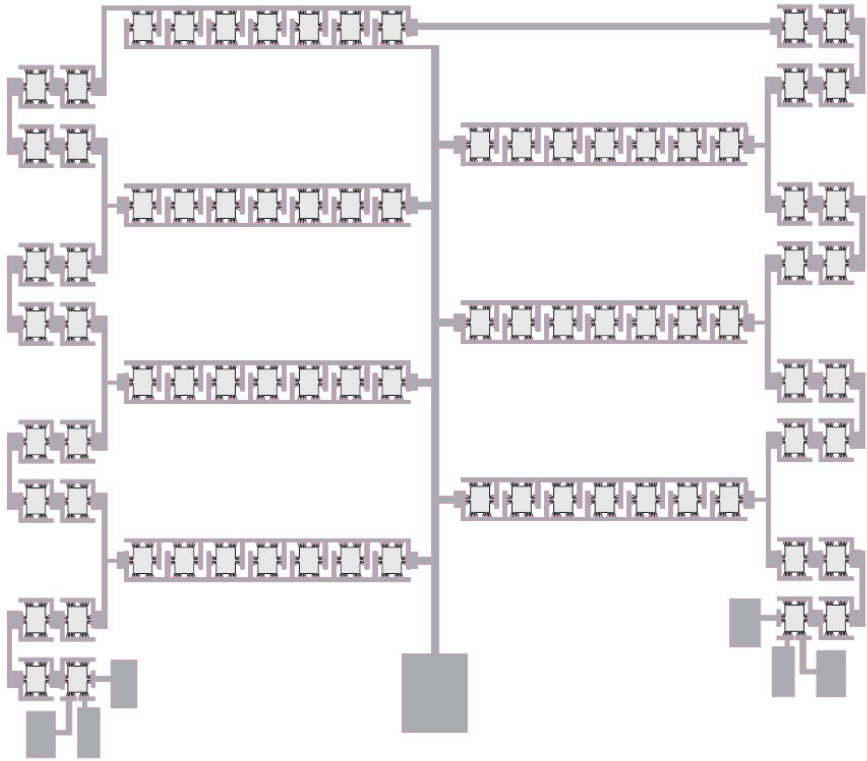
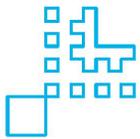
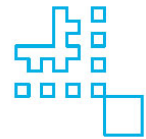

FIG. 2-SM. A design layout is shown for the following parameters: $M = 3$ and $\xi = 7$, $q_{M:i} = 4$, and $D_T = 81$. (Recall that $\xi$ represents the single QHR elements in parallel). The desired value was 1 P$\Omega$, and this arrangement reaches within approximately 8 % of that nominal value (that is, nearly 0.92 P$\Omega$).

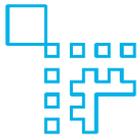
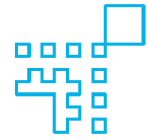
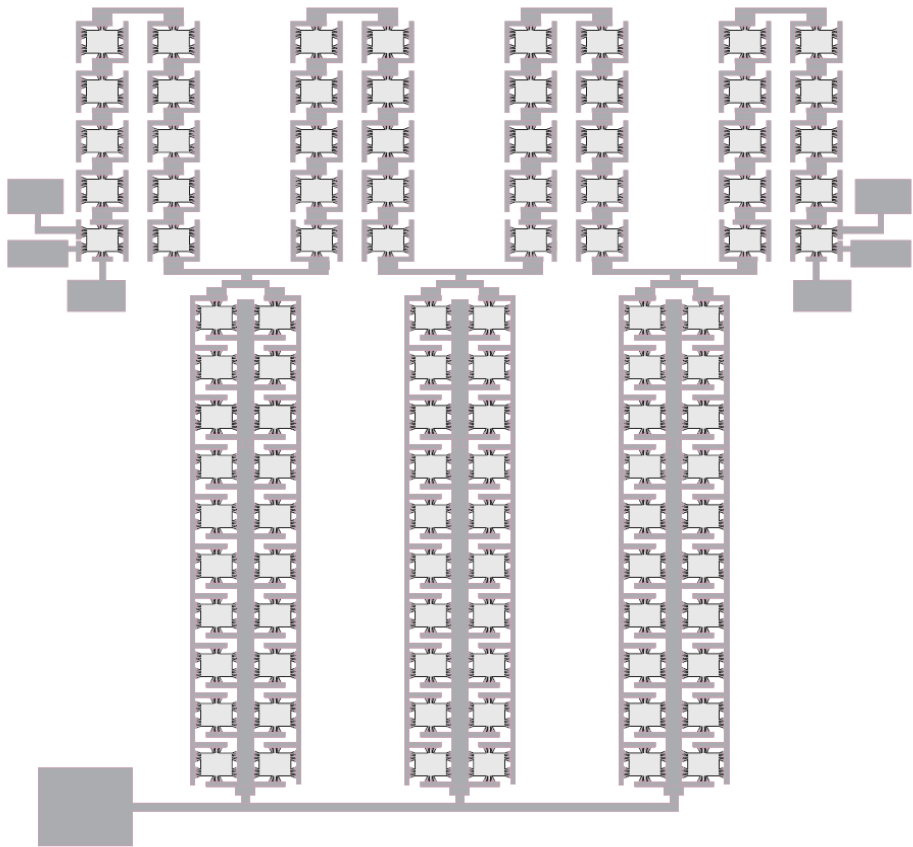
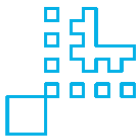
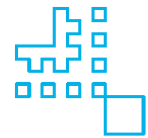

FIG. 3-SM. A design layout is shown for the following parameters: $M = 2$ and $\xi = 20$, $q_{M:i} = 10$, and $D_T = 100$. (Recall that $\xi$ represents the single QHR elements in parallel). The desired value was 1 T$\Omega$, and this arrangement reaches within approximately 5 % of that nominal value (that is, nearly 1.05 T$\Omega$).

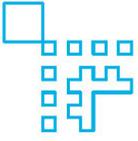
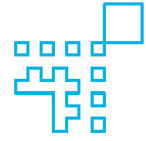

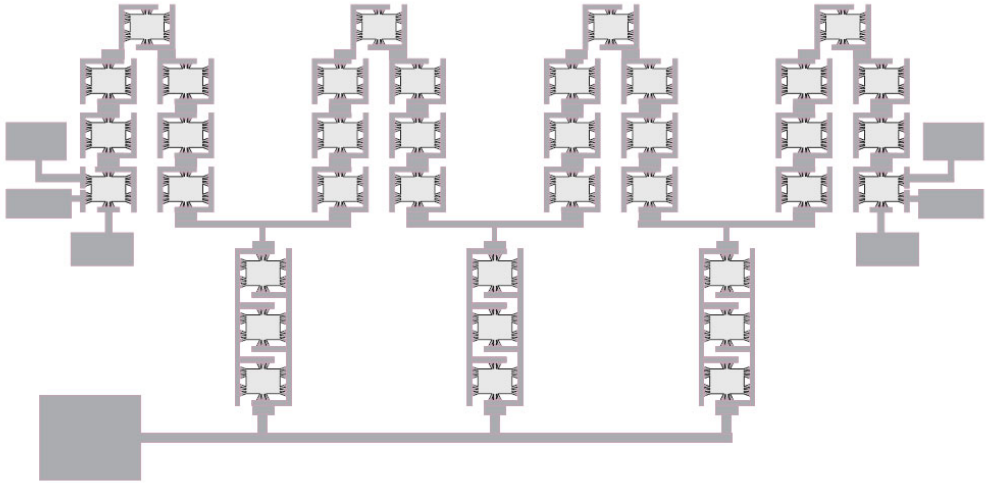

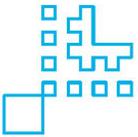
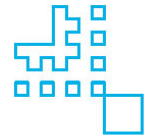

FIG. 4-SM. A design layout is shown for the following parameters: $M = 2$ and $\xi = 3$, $q_{M:i} = 7$, and $D_T = 37$. (Recall that $\xi$ represents the single QHR elements in parallel). The desired value was 1 GΩ, and this arrangement reaches within approximately 1 % of that nominal value (that is, nearly 1.01 GΩ).

## 4. QHARS Device Used in Experiments

The main QHARS device used was valued nominally at about 1 MΩ, shown in Fig. 5-SM.

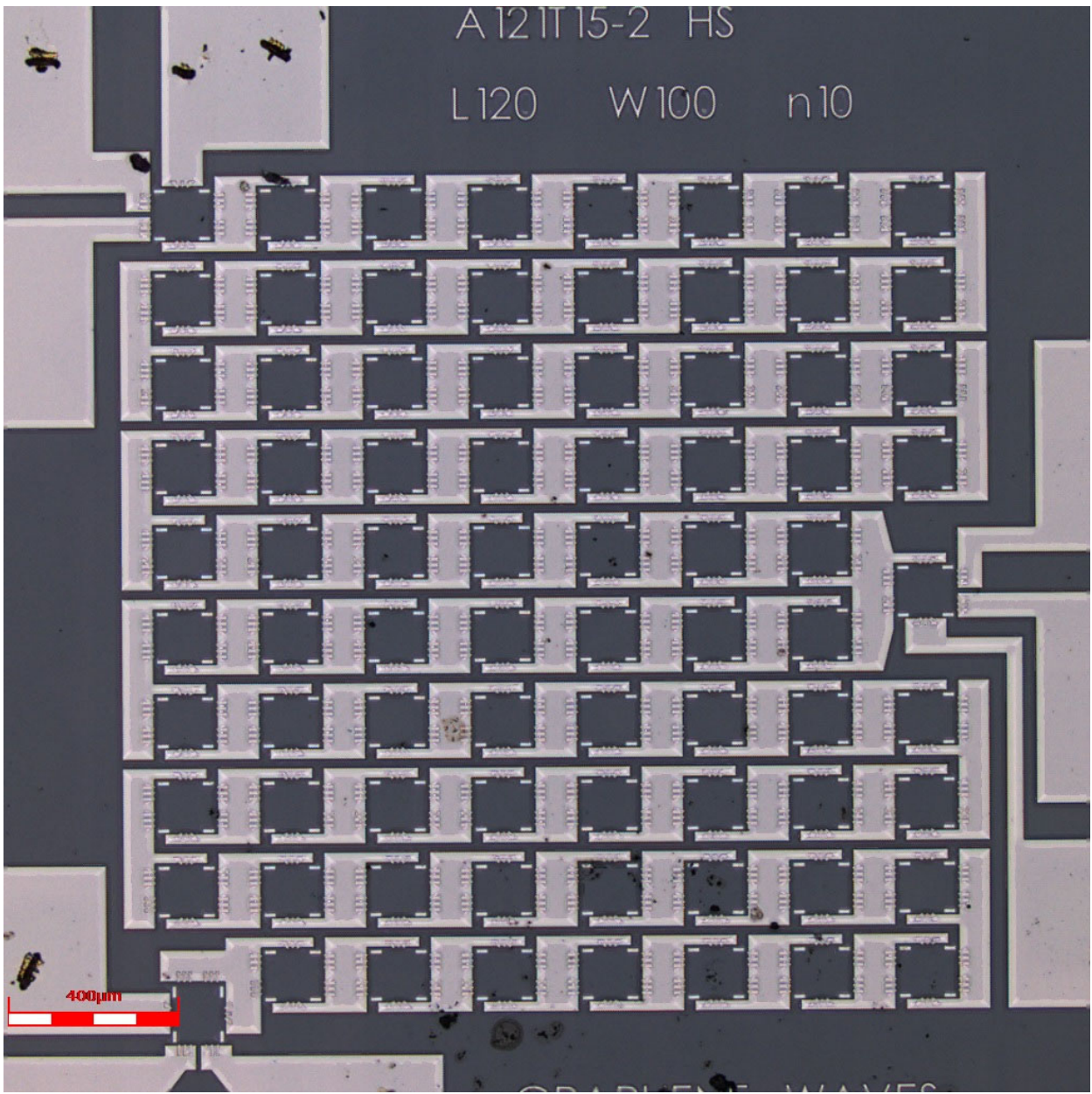

FIG. 5-SM. A QHARS device valued nominally at about 1 MΩ due to its 78 devices in series, with a middle element on the right side, center, as the grounded branch.